\documentclass[aps,prd,twocolumn,superscriptaddress]{revtex4}
\usepackage{epsfig,epsf}
\usepackage{amsmath}
\usepackage{amsthm}
\usepackage{amsfonts}
\usepackage{amssymb}
\usepackage{dsfont}
\usepackage{multirow}
\usepackage{appendix}
\usepackage{slashed}
\usepackage[active]{srcltx}
\usepackage{psfrag}
\usepackage{subfigure}
\newcommand{\be}{\begin{equation}}
\newcommand{\ee}{\end{equation}}
\newcommand{\ba}{\begin{eqnarray}}
\newcommand{\ea}{\end{eqnarray}}
\newcommand{\baa}{\begin{eqnarray*}}
\newcommand{\btab}{\begin{tabular}}
\newcommand{\etab}{\end{tabular}}
\newcommand{\eaa}{\end{eqnarray*}}

\def\inbar{\,\vrule height1.5ex width.4pt depth0pt}
\def\IC{\relax\hbox{$\inbar\kern-.3em{\rm C}$}}
\def\IZ{\relax{\hbox{\cmss Z\kern-.4em Z}}}
\def\IR{{\hbox{{\rm I}\kern-.2em\hbox{\rm R}}}}

\def\IP{{\hbox{{\rm I}\kern-.2em\hbox{\rm P}}}}
\def\II{\hbox{{1}\kern-.25em\hbox{l}}}

\begin{document}

\title{X(3872): propagating in a  dense medium}
\date{\today}
\author{K. Azizi$^{1,2}$, N. Er$^3$  \\
\textit{$^1$School of Physics, Institute for Research in Fundamental Sciences (IPM), P. O. Box 19395-5531, Tehran, Iran}\\
\textit{$^2$ Department of Physics, Do\v{g}u\c{s} University, Ac{\i}badem-Kad{\i}k\"oy, 34722 Istanbul, Turkey}\\
\textit{$^3$ Department of Physics, Abant \.{I}zzet Baysal University,
G\"olk\"oy Kamp\"us\"u, 14980 Bolu, Turkey}}

\begin{abstract}
In cold nuclear matter, the shifts in the mass and current-meson coupling as well the vector self energy of the exotic $X(3872)$ are calculated  using the diquark-antidiquark current within the framework of the in-medium two-point QCD sum rule.  At the rest frame of the medium, the three momentum of the considered particle is fixed to remove the contributions  of the particles with negative energy. In the calculations, we include the in-medium condensates of quark-quark, gluon-gluon and quark-gluon. It is observed that, the shift due to the nuclear matter is negative and is about $25\%$ when the saturation density is used. Such shift is considerably large and comparable with the nucleon' mass shift due to the nuclear medium. The negative shift in the current-meson coupling due to nuclear matter is approximately $10\%$. At the saturation density, the vector self energy of the exotic $X(3872)$ state is found to be $\Sigma_\upsilon=1.31$ GeV. It is shown that  the mass, current-meson coupling and vector self energy of $X(3872)$ strongly depend on the density of cold nuclear matter. 
\end{abstract}


\maketitle

\section{Introduction}

In recent years, the new hadronic states such as meson-meson, meson-baryon and baryon-baryon molecules, tetraquarks, pentaquarks, hybrids and glueballs have been in the focus of much attention. These exotic states can not be considered as the usual quark-antiquark or three-quark form of the well-known hadronic spectroscopy.  Both the quark model and QCD do not exclude the existence of such states. Hence, the investigation of such states can play essential role in  understanding of their structures and gaining useful knowledge on the perturbative and non-perturbative natures of QCD. 

 Among the exotic states, the tetraquark state $X(3872)$ is one of  the most interesting particles. Firstly, the Belle Collaboration \cite{Choi:2003ue} detected $X(3872)$ as  a narrow charmonium-like state produced in the exclusive decay
process  $B^{\pm} \rightarrow K^{\pm}\pi^+\pi^-J/\psi$, which decays into $\pi^+\pi^-J/\psi$, with a mass of  $m_X=3872.0 \pm
0.6$(stat) $\pm0.5$(syst)  MeV. It was confirmed later in some other experiments: CDF II Collaboration \cite{ Acosta:2003zx} via $\bar{p}p$  collisions with $m_X=3871.3 \pm 0.7$(stat) $\pm0.4$(syst)  MeV, D0 Collaboration  \cite{Abazov:2004kp} again in $\bar{p}p$  collisions  with the mass difference between the $X(3872)$ state and $J/\psi$  as  $774.9 \pm 3.1$ (stat)$ \pm 3.0$ (syst) MeV and  BABAR detector  \cite{Aubert:2004ns} at the PEP-II $e^+e^-$ asymmetric-energy storage ring with the mass  $3873.4 \pm 1.4 $ MeV. Belle Collaboration also assigned the quantum numbers of $X(3872)$ 
as $J^{PC}=1^{++}$ \cite{Abe:2005iya}. The decay width of  $X(3872)$ state was estimated as $\Gamma_{X(3872)}<1.2$ MeV \cite{Nielsen:2009uh,Choi:2011fc}.
 
 Following the discovery of $X(3872)$ both the theoretical and experimental researches on the non-conventional particles gained an acceleration. Such that, many tetraquark and pentaquark states have been discovered in the experiments and a  rush of theoretical  papers on the structures of the discovered and possible exotic particles appeared. Despite a lot of theoretical and experimental studies; the structure, nature and quark organisation of the newly founded states have not been understood exactly, yet. There are a lot of suggestions on the nature of the exotic  $X(3872)$  states, for instance, see  Refs. \cite{Ortega:2017hpw,Yu:2017bsj,Zhou:2017dwj,Wang:2015rcz,Ferretti:2015fba,Pena:2014pea,Guo:2014hqa,Wang:2013vex,Baru:2013rta,Wang:2013kva,Cui:2011be,Takizawa:2012hy,Coito:2012vf,Chen:2013pya,Ferretti:2014xqa,Karliner:2014lta,Meng:2014ota,Achasov:2015oia,Larionov:2015nea,Larionov:2015kxn,Maiani:2004vq,Bugg:2004rk,Li:2004sta,Seth:2004zb,Swanson:2003tb,Wang:2013daa,Braaten:2013poa,Chen:2016qju,Matheus:2006xi,Kang:2016jxw,Sundu:2016oda}. In Ref.  \cite{Maiani:2004vq}, for example, it is assumed as  a diquark-antidiquark states with hidden or open charm of the forms: $[cq][\bar{c}\bar{q}']$ and $[cq][\bar{s}\bar{q}']$ with $q,q'=u,d$.  In Ref. \cite{Bugg:2004rk}, it was suggested that it might be a $J^{PC}=1^{++}$ cusp due to the $D\bar{D}^*$ threshold. The author in Ref.  \cite{Li:2004sta} presented a study on the possibility of $X(3872)$ to be considered as a hybrid state of $c\bar{c}g$. In another study, Ref. \cite{Seth:2004zb}, it is proposed that $X(3872)$ can be a vector glueball mixed with neighbouring vector states of charmonium. Despite a lot of studies on the structure of $X(3872)$, it' nature remains unclear and needs to be investigated further.
  
Almost all studies on the properties of  $X(3872)$ and other exotic states have been performed in vacuum and there is almost no study on the in-medium properties of non-conventional particles, except Ref. \cite{Veliev:2017fpa} which investigates the thermal properties of $X(3872)$. Investigation of the in-medium properties of exotic states, besides the standard hadrons, can play crusial role in the analysing of the results of the heavy-ion collision experiments. Many experimental  collaborations, such as   PANDA, CBM at FAIR and NICA are planning to study the in-medium properties of hadrons including the  newly discovered non-conventional states \cite{Prencipe:2015cgg, Biswas:2015paa, fair, panda, Friman, Lutz:2009ff, Giacosa:2015nwa, nika}. For instance,  PANDA have focused on the search of the not-yet discovered glueballs and on the already discovered but not-yet understood $X, Y,Z$ states, \cite{Giacosa:2015nwa}. Therefore  investigation of exotic particles  in medium becomes an attractive and exciting subject. 

In the present study, we are going to evaluate the spectroscopic parameters such as the mass, current-meson coupling constant and vector self energy of the exotic $X(3872)$ at cold nuclear medium within the in-medium two-point QCD sum rule method and considering a diquark-antidiquark picture for this particle. We will look for the shifts on the quantities under consideration due to the nuclear matter at a saturation density.   We will discuss the variations of parameters of  $X(3872)$ with respect to the variations of the density, as well. 

This work is organised in the following way. In section II, the in-medium QCD sum rules for the mass and   current-meson coupling of the exotic $X(3872)$ (in what follows denoted as $X$) are obtained. In section III, the numerical analysis of the obtained in-medium sum rules are performed.  We investigate the variations of the quantities with respect to the variations of the density and calculate the shifts on the quantities under consideration due to the nuclear matter in this section. The last section includes our concluding remarks.

\section{In-medium QCD sum rules for the spectroscopic properties of the exotic $X(3872)$ }

In the framework of the QCD sum rules, for the calculation of the mass and current coupling constant of $X$, we start with the following in-medium two-point correlation function: 
\begin{equation}\label{ }
\Pi_{\mu\nu}(p)=i\int{d^4 xe^{ip\cdot x}\langle\psi_0|\mathcal{T}[\eta_{\mu}(x)\eta^{\dagger}_{\nu}(0)]|\psi_0\rangle},
\end{equation}
where $p$ is the four momentum of $X$, $|\psi_0\rangle$ is the parity and time-reversal symmetric ground state of the nuclear matter and $\eta_{\mu}(x)$ is the interpolating current of $X$. In  diquark-antidiquark model, the interpolating current of $X$  with the quantum numbers $J^{PC}=1^{++}$ is written as \cite{Matheus:2006xi}
\begin{eqnarray}\label{ }
 \eta_{\mu}(x)&=&\frac{i\epsilon_{abc}\epsilon_{dec}}{\sqrt{2}} \Big \{ \Big[q_{a}^T(x) C \gamma_5 c_b(x)\Big] \Big[\bar{q}_{d}(x) \gamma_{\mu} C\bar{c}^T_e(x) \Big] \nonumber \\
 &+& \Big[q_{a}^T(x) C \gamma_{\mu} c_b(x)\Big] \Big[\bar{q}_{d}(x) \gamma_5 C\bar{c}^T_e(x) \Big] \Big \},
\end{eqnarray}
where $a, b, c, d$ and $e$ are color indices, $q$ represents $u$ or $d$ light quark and $C$ is the charge conjugation matrix. To obtain the QCD sum rules for the mass and current coupling constant of $X$, the aforementioned correlation function is calculated in two different representations: hadronic (Had) and QCD with the help of operator product expansion (OPE). These two representations are equated through a dispersion relation to get the desired sum rules. To suppress the contributions of the higher states and continuum, a Borel transformation as well as continuum subtraction are applied to both sides of the obtained sum rules. 

\subsection{Hadronic Representation}
In the hadronic side, the correlation function is obtained in terms of the hadronic degrees of freedom. By performing integration over four-x, we get
\begin{equation} \label{had1}
\Pi^{Had}_{\mu\nu}(p)=- \frac{\langle\psi_0|\eta_{\mu}|X(p)\rangle \langle X(p)|\eta^{\dagger}_{\nu}|\psi_0\rangle}{p^{*2}-m^{*2}_X} + ... ,
\end{equation}
where $p^{*}$ is the in-medium momentum, $m^{*}_X$ is the modified mass of the X state due to the cold nuclear medium and $...$ is used for contributions of the higher resonances and  continuum. The decay constant or current-meson coupling is defined through the following matrix element in terms of the polarisation vector $\varepsilon_{\mu}$ of $X$ state:
\begin{equation} \label{had2}
\langle\psi_0|\eta_{\mu}|X(p)\rangle = f^{*}_X m^{*}_X \varepsilon_{\mu},
\end{equation}
where $f^{*}_X$ is the in-medium current coupling constant of the $X$ state.
After inserting Eq. (\ref{had2}) into Eq. (\ref{had1}), the hadronic side of the correlation function is obtained as
\begin{equation}
\Pi_{\mu\nu}^{Had}(p)=-\frac{m^{*2}_X f^{*2}_X}{p^{*2}-m^{*2}_X} \Big[-g_{\mu\nu} +\frac{p_{\mu}^{*} p^{*}_{\nu} }{m^{*2}_X} \Big],
\end{equation}
where the summation over spins of the Dirac spinors has been applied. To  proceed,  the in-medium momentum is represented in terms of the self energy $\Sigma_{\mu, \nu}$ as  $p_{\mu}^*=p_{\mu}-\Sigma_{\mu, v}$. The  self-energy   $\Sigma_{\mu, v}$ has an explicit form 
\begin{equation}\label{}
\Sigma_{\mu,v}=\Sigma_{v} u_{\mu} + \Sigma'_{v}p_{\mu},
\end{equation}
where $\Sigma_{v} $ is the vector self energy of $X$ and $u_{\mu}$ is the four velocity of the nuclear medium.  In mean-field approximation, the momentum independent scalar and vector self energies are taken to be real and the $ \Sigma'_{\nu}$ is identically zero \cite{Cohen:1994wm,Thomas:2007gx}. In this context, particles of any three-momentum  appear as stable quasi-particles  with self energies that are roughly linear in the density up to nuclear matter density \cite{Cohen:1994wm,Serot:1984ey}. The calculations  are performed in the rest frame of the nuclear medium, i.e. $u_{\mu}=(1,0)$ and at fixed
three-momentum of $X$ state, $|\vec{p}|=0.27~GeV$ (i.e. approximately the Fermi momentum). Note that in the vacuum, where the invariant functions depend only on $ p^2 $ the separation of $ p_0 $ and $\vec{p}  $ dependence is not necessary.  At finite density, however, we will keep the dispersion relations as integrals over $ p_0 $ with the three-momentum held fixed. This provides a clean identification of the intermediate quasiparticles. By this way, the contributions coming from the negative energy quasiparticles  is clearly separated, which lets us isolate  the contributions of the positive energy quasiparticles by adopting an appropriate weighting function.
After the replacement of the in-medium momentum in Eq. (5), the hadronic side becomes,
\begin{eqnarray}
\Pi_{\mu\nu}^{Had}(p)&=&-\frac{m^{*2}_X f^{*2}_X}{(p^2-2\Sigma_{\upsilon} p_0+\Sigma_{\upsilon}^2)-m^{*2}_X} \Big[-g_{\mu\nu}  \nonumber \\
&+&\frac{p_{\mu}p_{\nu} -\Sigma_{\upsilon}p_{\mu}u_{\nu} -\Sigma_{\upsilon}p_{\nu}u_{\mu}+\Sigma_{\upsilon}^2 u_{\mu}u_{\nu}}{m^{*2}_X} \Big], \nonumber \\
\end{eqnarray}
where $p_0=p\cdot u$ is the energy of the quasi-particle. Using the positive energy pole $E_p=\Sigma_{v}+\sqrt{|\vec{p}|^2+ m^{*2}_X}$ and the negative energy pole $\bar{E}_p=\Sigma_{v}-\sqrt{|\vec{p}|^2+ m^{*2}_X}$, we get
\begin{eqnarray}
\Pi_{\mu\nu}^{Had}(p)&=&-\frac{m^{*2}_X f^{*2}_X}{(p_0-E_p)(p_0-\bar{E}_p)} \Big[-g_{\mu\nu}  \nonumber \\
&+&\frac{p_{\mu}p_{\nu} -\Sigma_{\upsilon}p_{\mu}u_{\nu} -\Sigma_{\upsilon}p_{\nu}u_{\mu}+\Sigma_{\upsilon}^2 u_{\mu}u_{\nu}}{m^{*2}_X} \Big]. \nonumber \\
\end{eqnarray}
 In terms of spectral densities,  the hadronic correlation function can be written in an integral form  as
\begin{equation}\label{}
\Pi_{\mu\nu}^{Had}(p_0, \vec{p})=\frac{1}{2\pi i} \int_{-\infty}^{\infty} d\omega \frac{\Delta \rho_{\mu\nu}^{Had}(\omega, \vec{p})}{\omega-p_0},
\end{equation}
where the spectral densities $\Delta \rho_{\mu\nu}^{Had}(\omega, \vec{p})$ which are defined as
\begin{equation}\label{}
\Delta \rho_{\mu\nu}^{Had}(\omega, \vec{p})=Lim_{\epsilon \rightarrow 0^+} \Big[\Pi_{\mu\nu}^{Had}(\omega+i\epsilon,\vec{p})-\Pi_{\mu\nu}^{Had}(\omega-i\epsilon,\vec{p})\Big],
\end{equation}
are obtained in the following form:
\begin{eqnarray}\label{}
\Delta \rho_{\mu\nu}^{Had}(\omega, \vec{p})&=&\frac{m^{*2}_X f^{*2}_X}{2\sqrt{|\vec{p}|^2+ m^{*2}_X}}\Big[-g_{\mu\nu}  \nonumber \\
&+&\frac{p_{\mu}p_{\nu} -\Sigma_{\upsilon}p_{\mu}u_{\nu} -\Sigma_{\upsilon}p_{\nu}u_{\mu}+\Sigma_{\upsilon}^2 u_{\mu}u_{\nu}}{m^{*2}_X} \Big] \nonumber \\
&\times& \Big[ \delta(\omega - E_p) - \delta(\omega- \bar{E}_p) \Big] . \nonumber \\
\end{eqnarray}
It is necessary to remove the negative-energy pole contribution. For this purpose the correlation function is multiplied by the weight function $(\omega-\bar{E}_p)e^{\frac{-\omega^2}{M^2}}$ and the integral over $\omega$  is performed from $-\omega_0$ to $\omega_0$,
\begin{equation}\label{}
\Pi_{\mu\nu}^{Had}(p_0, \vec{p})=\int_{-\omega_0}^{\omega_0} d\omega  \Delta \rho_{\mu\nu}^{Had}(\omega, \vec{p}) (\omega-\bar{E}_p) e^{-\frac{\omega^2}{M^2}}.
\end{equation}
Here, $M^2$ is the Borel mass parameter and $\omega_0$ is the threshold parameter of the subtraction. These parameters will be fixed in numerical analyses. As a result of these calculations, the final form of the hadronic side of correlation function is obtained in terms of the corresponding structures as
\begin{eqnarray}\label{}
\Pi_{\mu\nu}^{Had}(p_0, \vec{p})&=& m^{*2}_X f^{*2}_X e^{-\frac{E_p^2}{M^2}}\Big[-g_{\mu\nu}  \nonumber \\
&+&\frac{p_{\mu}p_{\nu} -\Sigma_{\upsilon}p_{\mu}u_{\nu} -\Sigma_{\upsilon}p_{\nu} u_{\mu}+\Sigma_{\upsilon}^2 u_{\mu}u_{\nu}}{m^{*2}_X} \Big]. \nonumber \\
\end{eqnarray}

\subsection{OPE Representation}
The correlation function in QCD side is calculated in terms of QCD degrees of freedom in deep Euclidean region by the help of the operator product expansion. We can decompose the correlation function in QCD side in terms of different structures as
\begin{eqnarray}\label{}
\Pi_{\mu\nu}^{OPE}(p_0, \vec{p})&=& \Pi^{OPE}_1 g_{\mu\nu}  + \Pi^{OPE}_2 p_{\mu}p_{\nu}  + \Pi^{OPE}_3 p_{\mu}u_{\nu}  \nonumber \\
&+&\Pi^{OPE}_4 p_{\nu}u_{\mu} + \Pi^{OPE}_5 u_{\mu}u_{\nu}.
\end{eqnarray}
The coefficient of each structure  can be represented in terms of the corresponding spectral density  as
\begin{equation}\label{}
\Pi^{OPE}_{i}(p_0, \vec{p})= \frac{1}{2\pi i}\int_{-\infty}^{\infty} d\omega \frac{\Delta \rho^{OPE}_{i}(\omega, \vec{p})}{\omega-p_0},
\end{equation}
where $\Delta \rho^{OPE}_{i}(\omega, \vec{p})$ are the imaginary parts of the correlation functions $\Pi^{OPE}_{i}(p_0, \vec{p})$. For the calculation of spectral densities, the explicit form of the interpolating current is substituted  into the correlation function in Eq. (1) and all quark pairs are contracted by applying the Wick's theorem. As a result, we get
\begin{eqnarray}\label{}
&&\Pi_{\mu\nu}^{OPE}(p)= -\frac{i}{2}\varepsilon_{abc} \varepsilon_{a'b'c'} \varepsilon_{dec} \varepsilon_{d'e'c'} \nonumber \\
&\times& \int d^4 x e^{ipx} \langle\psi_0|\Big\{Tr\Big[\gamma_5 \tilde{S}_q^{aa'} (x)\gamma_5 S_c^{bb'}(x)\Big] \nonumber\\
&\times& Tr\Big[\gamma_{\mu} \tilde{S}_c^{ee'} (-x)\gamma_{\nu} S_q^{dd'}(-x)\Big] + Tr\Big[[\gamma_{\mu} \tilde{S}_c^{ee'} (-x) \nonumber\\
&\times& \gamma_5 S_q^{dd'}(-x)\Big]Tr\Big[\gamma_{\nu} \tilde{S}_q^{aa'} (x)\gamma_5 S_c^{bb'}(x)\Big] \nonumber \\
&+& Tr\Big[\gamma_5 \tilde{S}_q^{aa'} (x)\gamma_{\mu} S_c^{bb'}(x)\Big] Tr\Big[\gamma_5 \tilde{S}_c^{ee'} (-x)\gamma_{\nu} S_q^{dd'}(-x)\Big] \nonumber \\
&+& Tr\Big[\gamma_{\nu}  \tilde{S}_q^{aa'} (x)\gamma_{\mu} S_c^{bb'}(x)\Big] \nonumber \\
&\times&Tr\Big[\gamma_5 \tilde{S}_c^{ee'} (-x)\gamma_5 S_q^{dd'}(-x)\Big]  \Big\}|\psi_0\rangle,\nonumber \\
\end{eqnarray}
in terms of the light and heavy quark propagators. The abbreviation $\tilde{S}_{q(c)}= C S_{q(c)}^{T}C$ has been used in the above equation.
The light and heavy quark propagators in  coordinate space and in  the fixed point gauge are written in $m_q\rightarrow 0$ limit as
\begin{eqnarray}
S_q^{ij}(x)&=&
\frac{i}{2\pi^2}\delta^{ij}\frac{1}{(x^2)^2}\not\!x
 + \chi^i_q(x)\bar{\chi}^j_q(0) \nonumber \\
&-&\frac{ig_s}{32\pi^2}F_{\mu\nu}^{ij}(0)\frac{1}{x^2}[\not\!x\sigma^{\mu\nu}+\sigma^{\mu\nu}\not\!x] +\cdots \, ,
\end{eqnarray}
and
\begin{eqnarray}
S_c^{ij}(x)&=&\frac{i}{(2\pi)^4}\int d^4k e^{-ik \cdot x} \left\{\frac{\delta_{ij}}{\!\not\!{k}-m_c}\right.\nonumber\\
&&\left.-\frac{g_sF_{\mu\nu}^{ij}(0)}{4}\frac{\sigma_{\mu\nu}(\!\not\!{k}+m_c)+(\!\not\!{k}+m_c)
\sigma_{\mu\nu}}{(k^2-m_c^2)^2}\right.\nonumber\\
&&\left.+\frac{\pi^2}{3} \langle \frac{\alpha_sGG}{\pi}\rangle\delta_{ij}m_c \frac{k^2+m_c\!\not\!{k}}{(k^2-m_c^2)^4}+\cdots\right\} \, , \nonumber \\
\end{eqnarray}
where $\chi^i_q$ and $\bar{\chi}^j_q$ are the Grassmann background quark fields. We use the short-hand notation
\begin{equation}
\label{ }
F_{\mu\nu}^{ij}=F_{\mu\nu}^{A}t^{ij,A}, ~~~~~A=1,2, ...,8,
\end{equation}
where, $F_{\mu\nu}^A$ are classical background gluon fields, and $t^{ij,A}=\frac{\lambda ^{ij,A}}{2}$ with $
\lambda ^{ij, A}$ being  the standard Gell-Mann matrices. Using Eqs. (17) and (18)  in Eq. (16), we get  the products
of the Grassmann background quark fields and classical background
gluon fields which correspond to the ground-state matrix elements
of the corresponding quark and gluon operators \cite{Cohen:1994wm, Azizi:2014yea},
\begin{eqnarray}\label{}
\chi_{a\alpha}^{q}(x)\bar{\chi}_{b\beta}^{q}(0)&=&\langle
q_{a\alpha}(x)\bar{q}_{ b\beta}(0)\rangle_{\rho_N},
 \nonumber \\
F_{\kappa\lambda}^{A}F_{\mu\nu}^{B}&=&\langle
G_{\kappa\lambda}^{A}G_{\mu\nu}^{B}\rangle_{\rho_N}, \nonumber \\
\chi_{a\alpha}^{q}\bar{\chi}_{b\beta}^{q}F_{\mu\nu}^{A}&=&\langle
q_{a\alpha}\bar{q}_{ b\beta}G_{\mu\nu}^{A}\rangle_{\rho_N},
 \nonumber \\
 \mbox{and }
 \nonumber \\
\chi_{a\alpha}^{q}\bar{\chi}_{b\beta}^{q}\chi_{c\gamma}^{q}\bar
{\chi}_{d\delta}^{q}&=&\langle q_{a\alpha}\bar{q}_{b\beta}
q_{c\gamma}\bar{q}_{d\delta}\rangle_{\rho_N},
\end{eqnarray}
where $\rho_N$ is the nuclear matter density.

To proceed; the quark, gluon and mixed
condensates are needed to be defined in nuclear matter. The matrix element $\langle
q_{a\alpha}(x)\bar{q}_{b\beta}(0)\rangle_{\rho_N}$ can be parameterized 
 as \cite{Cohen:1994wm}
\begin{eqnarray} \label{ }
\langle
q_{a\alpha}(x)\bar{q}_{b\beta}(0)\rangle_{\rho_N}&=&-\frac{\delta_{ab}}{12}
\Bigg
[\Bigg(\langle\bar{q}q\rangle_{\rho_N}+x^{\mu}\langle\bar{q}D_{\mu}q\rangle_{
\rho_N}
 \nonumber \\
&+&\frac{1}{2}x^{\mu}x^{\nu}\langle\bar{q}D_{\mu}D_{\nu}q\rangle_{\rho_N}
+...\Bigg)\delta_{\alpha\beta}\nonumber \\
&&+\Bigg(\langle\bar{q}\gamma_{\lambda}q\rangle_{\rho_N}+x^{\mu}\langle\bar{q}
\gamma_{\lambda}D_{\mu} q\rangle_{\rho_N}
 \nonumber \\
&+&\frac{1}{2}x^{\mu}x^{\nu}\langle\bar{q}\gamma_{\lambda}D_{\mu}D_{\nu}
q\rangle_{\rho_N}
+...\Bigg)\gamma^{\lambda}_{\alpha\beta} \Bigg].\nonumber \\
\end{eqnarray}
The quark-gluon mixed condensate can also be defined  as
\begin{eqnarray} \label{ }
&&\langle
g_{s}q_{a\alpha}\bar{q}_{b\beta}G_{\mu\nu}^{A}\rangle_{\rho_N}\nonumber \\&&=-\frac{t_{
ab}^{A }}{96}\Bigg\{\langle g_{s}\bar{q}\sigma\cdot
Gq\rangle_{\rho_N}
\Bigg[\sigma_{\mu\nu}+i(u_{\mu}\gamma_{\nu}-u_{\nu}\gamma_{\mu
})
\!\not\! {u}\Bigg]_{\alpha\beta} \nonumber \\
&&+\langle g_{s}\bar{q}\!\not\! {u}\sigma\cdot
Gq\rangle_{\rho_N}
 \Bigg[\sigma_{\mu\nu}\!\not\!
{u}+i(u_{\mu}\gamma_{\nu}-u_{\nu}\gamma_{\mu}
)\Bigg]_{\alpha\beta} \nonumber \\
&&-4\Bigg(\langle\bar{q}u\cdot D u\cdot D
q\rangle_{\rho_N}
+im_{q}\langle\bar{q}
\!\not\! {u}u\cdot D q\rangle_{\rho_N}\Bigg) \nonumber \\
&&\times\Bigg[\sigma_{\mu\nu}+2i(u_{\mu}\gamma_{\nu}-u_{\nu}\gamma_{\mu}
)\!\not\! {u}\Bigg]_{\alpha\beta}\Bigg\},
\nonumber \\
\end{eqnarray}
where 
$D_\mu=\frac{1}{2}(\gamma_\mu \!\not\!{D}+\!\not\!{D}\gamma_\mu)$ is the four-derivative.
The matrix element of the four-dimension gluon condensate can also be parametrized as 
\begin{eqnarray}
 \langle
G_{\kappa\lambda}^{A}G_{\mu\nu}^{B}\rangle_{\rho_N}&=&\frac{\delta^{AB}}{96}
\Bigg[ \langle
G^{2}\rangle_{\rho_N}(g_{\kappa\mu}g_{\lambda\nu}-g_{\kappa\nu}g_{\lambda\mu}
)
\nonumber \\
&+&O(\langle \textbf{E}^{2}+\textbf{B}^{2}\rangle_{\rho_N})\Bigg],
\end{eqnarray}
where the last term in this equation has a negligible contribution, so it can be ignored safely. The different operators in Eqs. (20-22) 
are defined as \cite{Cohen:1994wm,Jin:1992id}:
\begin{eqnarray} \label{ }
\langle\bar{q}\gamma_{\mu}q\rangle_{\rho_N}&=&\langle\bar{q}\!\not\!{u}q\rangle_
{\rho_N}
u_{\mu} ,\nonumber \\
\langle\bar{q}D_{\mu}q\rangle_{\rho_N}&=&\langle\bar{q}u\cdot D
q\rangle_{\rho_N}
u_{\mu}=-im_{q}\langle\bar{q}\!\not\!{u}q\rangle_{\rho_N}
u_{\mu}  ,\nonumber \\
\langle\bar{q}\gamma_{\mu}D_{\nu}q\rangle_{\rho_N}&=&\frac{4}{3}\langle\bar{q}
\!\not\! {u}u\cdot D
q\rangle_{\rho_N}(u_{\mu}u_{\nu}-\frac{1}{4}g_{\mu\nu})\nonumber \\
&+&\frac{i}{3}m_{q}
\langle\bar{q}q\rangle_{\rho_N}(u_{\mu}u_{\nu}-g_{\mu\nu}),
\nonumber \\
\langle\bar{q}D_{\mu}D_{\nu}q\rangle_{\rho_N}&=&\frac{4}{3}\langle\bar{q}
u\cdot D u\cdot D
q\rangle_{\rho_N}(u_{\mu}u_{\nu}-\frac{1}{4}g_{\mu\nu})\nonumber \\
&-&\frac{1}{6} \langle
g_{s}\bar{q}\sigma\cdot Gq\rangle_{\rho_N}(u_{\mu}u_{\nu}-g_{\mu\nu}) , \nonumber \\
\langle\bar{q}\gamma_{\lambda}D_{\mu}D_{\nu}q\rangle_{\rho_N}&=&2\langle\bar{q}
\!\not\! {u}u\cdot D u\cdot D q\rangle_{\rho_N}
\nonumber \\
&&\Bigg[u_{\lambda}u_{\mu}u_{\nu} -\frac{1}{6}
(u_{\lambda}g_{\mu\nu}+u_{\mu}g_{\lambda\nu}+u_{\nu}g_{\lambda\mu})\Bigg]\nonumber\\
&&-\frac{1}{6} \langle g_{s}\bar{q}\!\not\! {u}\sigma\cdot
Gq\rangle_{\rho_N}(u_{\lambda}u_{\mu}u_{\nu}-u_{\lambda}g_{\mu\nu}),\nonumber
\\
\end{eqnarray}
where, in their derivations,  the  equation of motion has been used and the terms
$\textit{O}(m^2_q)$  have been neglected due to
 their ignorable  contributions \cite{Cohen:1994wm}.

After using the above in-medium operators, the correlation function in the coordinate space is derived. In order to transfer the calculations to momentum space, we use
\begin{eqnarray}
\label{ }
\frac{1}{(x^2)^m}&=&\int \frac{d^Dk }{(2\pi)^D}e^{-ik \cdot x}i(-1)^{m+1}2^{D-2m}\pi^{D/2} \nonumber \\
&\times& \frac{\Gamma[D/2-m]}{\Gamma[m]}\Big(-\frac{1}{k^2}\Big)^{D/2-m},
\end{eqnarray}
and replace  $x_{\nu}$ by $-i\frac{\partial}{\partial p_{\nu}}$. The following formula is used to perform the resultant  four-integrals:
\begin{equation}
\label{ }
\int d^4 \ell\frac{(\ell^2)^m}{(\ell^2+\Delta)^n}=\frac{i\pi^2 (-1)^{m-n} \Gamma[m+2]\Gamma[n-m-2]}{\Gamma[2]\Gamma[n] (-\Delta)^{n-m-2}}.
\end{equation} 
By the help of the following relation, we extract the imaginary parts corresponding to different structures:
\begin{equation}
\label{ }
\Gamma\Big[\frac{D}{2}-n\Big]\Big(-\frac{1}{\Delta}\Big)^{D/2-n}=\frac{(-1)^{n-1}}{(n-2)!}(-\Delta)^{n-2}ln[-\Delta].
\end{equation} 
To get rid of the contributions of the negative energy particles, the expressions in the OPE side are multiplied by the weight function $(w-\bar{E}_p)e^{-\frac{w^2}{M^2}}$, as well and the following integrals are performed over $ w $: 
\begin{eqnarray} \label{Pi1}
\Pi^{OPE}_i(w_0,\vec{p})=\int^{w_0}_{-w_0} dw
\Delta\rho^{OPE}_i(w,\vec{p})(w-\bar{E}_p)e^{-\frac{w^2}{M^2}}. \nonumber \\
\end{eqnarray} \label{Pi1}
 The obtained results  are functions of $ w_0 $ and $ M^2 $. Applying the continuum subtraction procedure (for details see for instance \cite{Agaev:2016srl}) with the aim of more suppression of the contributions of the higher states and continuum we obtain the following  representation  in the Borel scheme: 
\begin{equation}
\label{ }
\Pi_{i}^{OPE}(s_0^{*},M^2)=\int_{4m_c^2}^{s_0^{*}} ds \rho_i^{OPE} (s) e^{\frac{-s}{M^2}},
\end{equation}
where we have used $w_0=\sqrt{s_0^*}$ with $s_0^*$ being the continuum threshold in nuclear matter.  Here,   $\rho_i^{OPE} (s)$ are the new spectral densities  and they are given in terms of the perturbative (pert), two-quark condensate, two-gluon  and mixed quark-gluon  condensates parts as
\begin{equation}
\label{ }
\rho^{OPE} (s)=\rho^{pert} (s)+\rho^{\langle q\bar{q}\rangle_{\rho_N} } (s)+\rho^{\langle GG \rangle_{\rho_N}} (s)+\rho^{\langle qG\bar{q} \rangle_{\rho_N}} (s).
\end{equation}
As examples , we present the explicit forms of the above spectral densities corresponding to the structure $g_{\mu\nu}$ in Appendix.

After getting the hadronic and the OPE sides of the correlation function and matching the coefficients of the selected structures form  both sides, the following sum rules are obtained 
\begin{eqnarray}
-m^2_{X}f^{*2}_X e^{-\frac{E_p^2}{M^2}}& = & \Pi^{OPE}_{1}, \nonumber  \\
f^{*2}_X e^{-\frac{E_p^2}{M^2}}& = & \Pi^{OPE}_{2}, \nonumber  \\
-\Sigma_{\upsilon}f^{*2}_X e^{-\frac{E_p^2}{M^2}}& = & \Pi^{OPE}_{3}, \nonumber  \\
-\Sigma_{\upsilon}f^{*2}_X e^{-\frac{E_p^2}{M^2}}& = & \Pi^{OPE}_{4}, \nonumber  \\
\Sigma^2_{\nu}f^{*2}_X e^{-\frac{E_p^2}{M^2}}& = & \Pi^{OPE}_{5},
\end{eqnarray}
which will be used in our numerical calculations.

\section{ Numerical  Analyses }
For the numerical  analyses of the physical quantities under consideration in nuclear matter and vacuum, we need the value of the saturated nuclear matter density and the in-medium and vacuum operators, whose numerical values are presented in Table 1.
\begin{table}[ht!]
\centering
\begin{tabular}{ |c | c| c|}
\hline \hline
Input  &  Value & Unit  \\ \hline
$ \rho_{N}^{sat} $   &  $0.11^3 $   & $GeV^3$    \\  \hline
$\langle q^{\dag}q\rangle_{\rho_N} $ &  $\frac{3}{2}\rho_{N}$ & $GeV^3$    \\  \hline
$ \langle\bar{q}q\rangle_{0} $ &  $ (-0.241)^3  $  & $GeV^3$    \\  \hline
$ m_q$ &  $ 0.00345  $  & $GeV$    \\  \hline
$ \sigma_N$ &  $ 0.059  $  & $GeV$    \\  \hline
$ \langle\bar{q}q\rangle_{\rho_N} $&  $\langle\bar{q}q\rangle_{0}+ \frac{\sigma_N}{2 m_q} \rho_N  $ & $GeV^3$    \\  \hline
$ \langle q^{\dag}iD_0q\rangle_{\rho_N}$&  $0.18 \rho_N  $ & $GeV^4$    \\  \hline
$ \Big\langle\frac{\alpha_s}{\pi}G^2\Big\rangle_{0}$&  $(0.33 \pm 0.04)^4\rho_N  $ & $GeV^4$    \\  \hline
$\Big\langle \frac{\alpha_s}{\pi}G^2\Big\rangle_{\rho_N}$& $\Big\langle \frac{\alpha_s}{\pi}G^2\Big\rangle_{0}-0.65~GeV \rho_N  $ & $GeV^4$    \\  \hline
$ \langle\bar{q}g_s\sigma Gq \rangle_{0}$ & $ 0.8 \langle\bar{q}q\rangle_{0}  $ & $GeV^5$    \\  \hline
$ \langle\bar{q}g_s\sigma Gq \rangle_{\rho_N}$ & $ \langle\bar{q}g_s\sigma Gq \rangle_{0} + 3 GeV^2 \rho_N$  & $GeV^5$    \\  \hline
$ \langle\bar{q}iD_0 iD_0q\rangle_{\rho_N} $&  $0.3 \rho_N -\frac{1}{8} \langle\bar{q}g_s\sigma Gq \rangle_{\rho_N} $ & $GeV^5$    \\  \hline
\hline \hline
\end{tabular}
\caption{Numerical inputs \cite{Cohen:1991nk,Belyaev:1982cd, Ioffe:2005ym}.}
\end{table}

Besides these input parameters, the obtained QCD sum rules are also functions of two auxiliary  parameters: the Borel mass parameter $M^2$ and the continuum threshold $s^*_0$. The working intervals for these auxiliary parameters are found based on the standard prescriptions of the method used. The standard criteria for the calculation of working intervals for these parameters are weak dependence of the results on these parameters, achieving to the maximum pole contributions (PC), convergence of the series of the OPE and exceeding of the perturbative parts over the non-perturbative contributions. As a result, we get 
\begin{equation}
\label{ }
3 ~ \textrm{GeV}^2 \leqslant M^2 \leqslant 5 ~ \textrm{GeV}^2,
\end{equation}
for $M^2$ and
\begin{equation}
\label{ }
4.1^2 ~ \textrm{GeV}^2 \leqslant s^*_0 \leqslant 4.3^2 ~ \textrm{GeV}^2,
\end{equation}
for $s^*_0$.
Considering these intervals of the auxiliary parameters we plot the pole contribution versus $ M^2 $ in figure 1 at different fixed values of the in-medium continuum threshold.
\begin{figure}[h!]
\label{fig1}
\centering
\begin{tabular}{ccc}
\epsfig{file=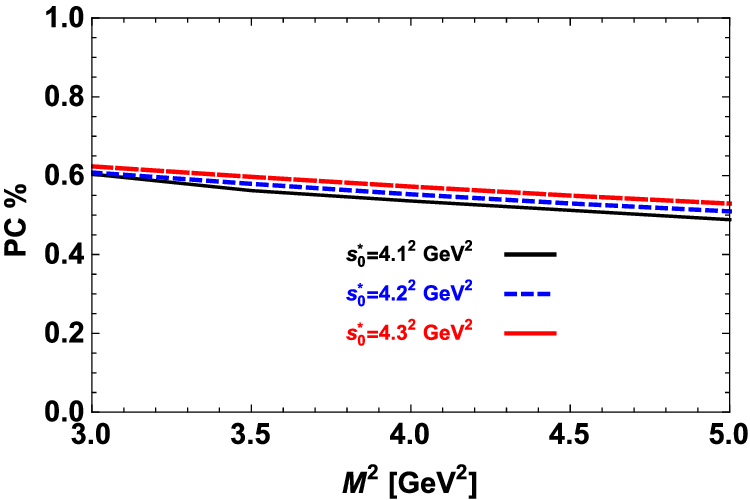,width=0.90\linewidth,clip=} \\ 
\end{tabular}
\caption{PC versus $M^2$  at different  fixed values of the in-medium continuum threshold.}
\end{figure}
From this figure we extract the average value of PC which is $ 56\% $ of the total contribution. This is a good amount of pole contribution in a tetraquark channel.

Now, to see how the results vary with respect to the Borel mass $M^2$, in figure 2, we plot the ratio of the mass in nuclear matter to the mass in vacuum, $m_X^*/m_X$, (top-panel), the ratio of the vector self energy to vacuum mass, $\Sigma_\upsilon/m_X$ (mid-panel) and the ratio of the in-medium current-meson coupling to it' vacuum value, $f_X^*/f_X$, (bottom-panel) versus $M^2$ for three different values of continuum threshold and at the saturated nuclear matter density, $\rho^{sat}=0.11^3$ GeV$^3$.  
As is seen, the results very weakly depend on the variations of the Borel parameter $M^2$. We also see very small variations of the results when we change the value of the in-medium threshold in its working region. Obtained from our analyses, we find an average $\%25$ negative shift in the mass of $X$ particle due to the cold nuclear medium when the saturation density is used. The shift in the meson-current coupling is about $\%10$ and is negative. Such shifts in the mass and residue of $X$ are comparable with those of the nucleon obtained in Ref. \cite{Azizi:2014yea}. These considerable shifts in the values of the parameters of $X$ can be attributed to its up/down quark content which strongly interacts with the nucleons in the medium. We find the average value of the vector self energy of $X$ particle as $\Sigma_{\upsilon}=1.31$ GeV at saturation density, $\rho_{N}^{sat} =0.11^3$.

It  is also possible to investigate the effects of nuclear matter density to the shifts on the mass, vector self energy and  the current-meson coupling. For this purpose, in figure 3,  we plot $m_X^{*}/m_X$ (top-panel),  $\Sigma_{\upsilon}/m_X$ (mid-panel) and $f_X^{*}/f_X$ (bottom-panel)  versus $\rho_N/\rho_N^{sat}$  at the average value of the Borel parameter and at fixed values of continuum thresholds. As is seen from these figures, the ratios $m_X^{*}/m_X$,  $\Sigma_{\upsilon}/m_X$ and $f_X^{*}/f_X$ strongly depend on the density. It is worth noting that the dependencies of these ratios on the nuclear matter density are roughly linear. 
\begin{figure}[h!]
\label{fig1}
\centering
\begin{tabular}{ccc}
\epsfig{file=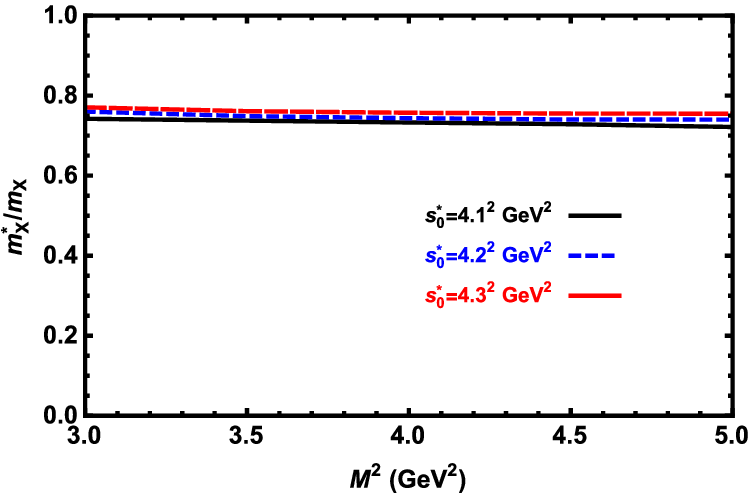,width=0.90\linewidth,clip=} \\
\epsfig{file=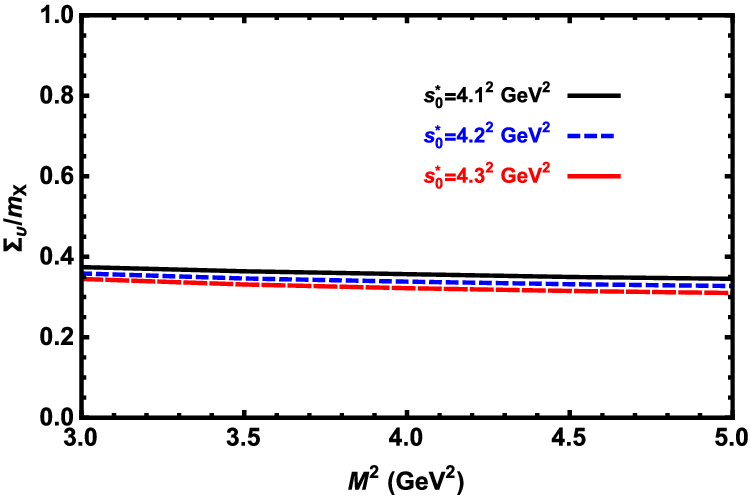,width=0.90\linewidth,clip=}  \\
\epsfig{file=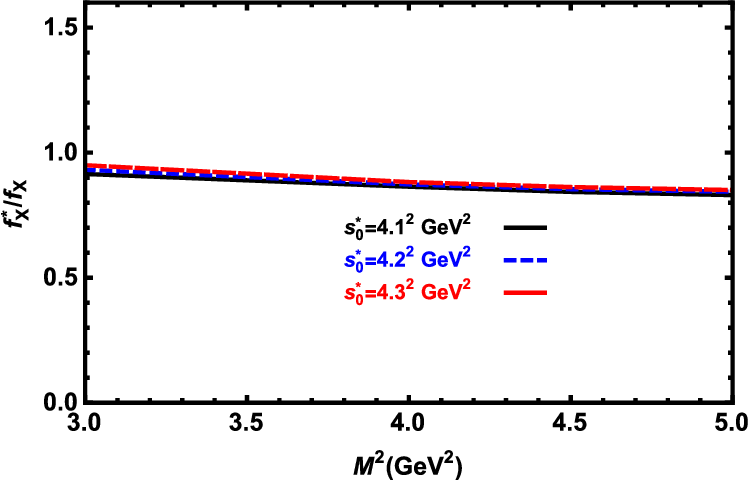,width=0.90\linewidth,clip=}  
\end{tabular}
\caption{$m_X^{*}/m_X$,  $\Sigma_{\upsilon}/m_X$ and $f_X^{*}/f_X$ as  functions of M$^2$ at the saturated nuclear matter density, $\rho_N^{sat}=0.11^3$ GeV$^3$ and at fixed values of the continuum threshold.}
\end{figure}

\begin{figure}[h!]
\label{fig1}
\centering
\begin{tabular}{ccc}
\epsfig{file=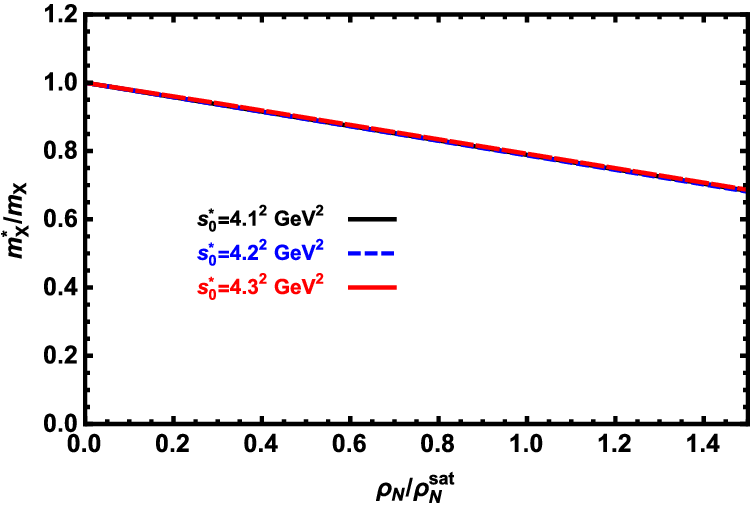,width=0.90\linewidth,clip=} \\
\epsfig{file=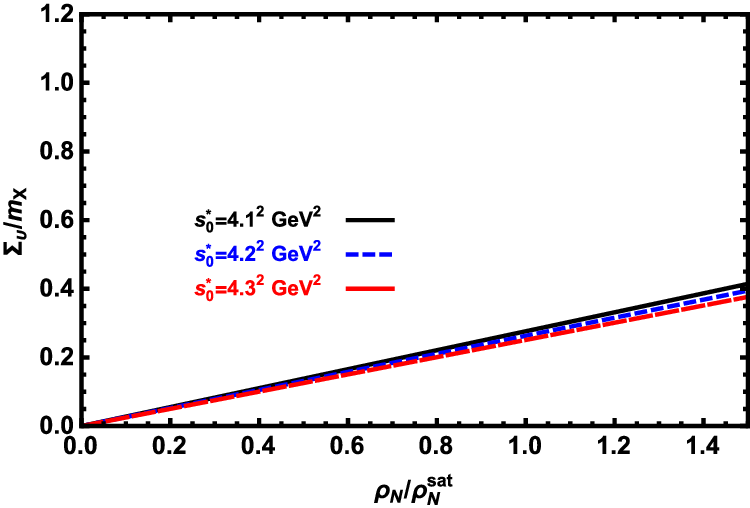,width=0.90\linewidth,clip=}  \\
\epsfig{file=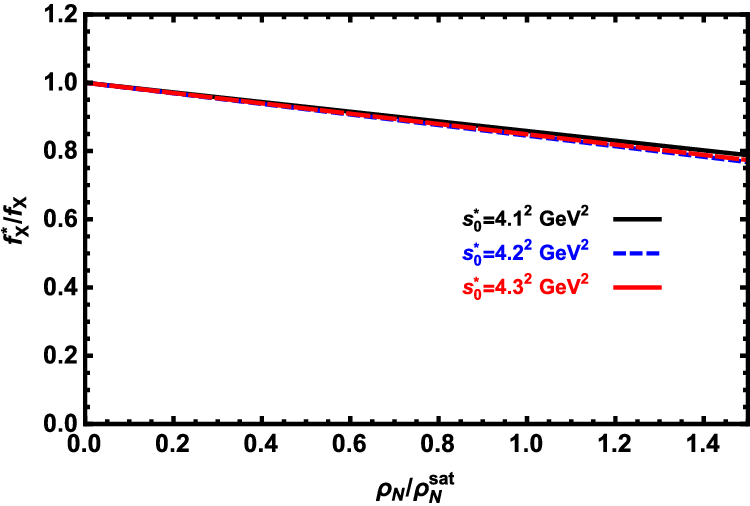,width=0.90\linewidth,clip=}  
\end{tabular}
\caption{$m_X^{*}/m_X$,  $\Sigma_{\upsilon}/m_X$ and $f_X^{*}/f_X$ versus $\rho/\rho_N^{sat}$  at different values of the threshold parameter $s_0^{*}$ and average value of $M^2$.}
\end{figure}

\section{Conclusion}
In this study, we used the interpolating current in a diquark-antidiquark form to investigate the spectroscopic parameters of the famous $X(3872)$ tetraquark in the cold nuclear matter. This is the first attempt to calculate the properties of  non-conventional particles in nuclear matter. The properties of  standard hadrons in nuclear medium were previously investigated extensively in the literature (see for instance  Refs. \cite{Azizi:2016dmr,Hosaka:2016ypm,Wang:2012xk,Wang:2011yj,Wang:2011hta,Cohen:1994wm,Furnstahl:1995nd,Drukarev:2004fn,Drukarev:2001wd,Koike:1996ga,Hayashigaki:2000es,Ohtani:2016pyk,Suzuki:2015est} and references theirin). We derived the in-medium sum rules to evaluate the shifts in the mass and current-meson coupling of the $X(3872)$ as well as found its vector self energy. The obtained results reveal that the mass of $X(3872)$ is considerably affected by the nuclear matter. We found a negative $25\%$ shift in its mass when the saturation density was used. The shift in the current-meson coupling  is also negative and it is roughly $10\%$ of its vacuum value. We found the vector self energy of  $X(3872)$ at the saturated density to be $\Sigma_{\upsilon}=1.31$ GeV. The order of shifts in the mass and residue of this particle is comparable with these of the nucleon investigated in Ref. \cite{Azizi:2014yea}.

We also investigated the dependence of the ratios $m_X^{*}/m_X$,  $\Sigma_{\upsilon}/m_X$ and $f_X^{*}/f_X$ on the nuclear matter density. We observed that, although the dependencies are linear  the results strongly depend on the nuclear matter density. 

The results obtained in the present work may be used in analyses of the heavy ion collision experiments as well as those which are aiming to investigate the properties of the standard and non-conventional hadrons in nuclear medium such as PANDA Collaboration at FAIR. 

Any experimental results on the in-medium properties of $X(3872)$ and comparison of those with the results of the present study can increase our knowledge of the exotic states and help us gain useful information on the not well-known structures of the exotic states, specially the newly discovered tetraquarks.

K.~A. thanks Do\v{g}u\c{s} University for the partial financial support
through the grant BAP 2015-16-D1-B04.

\appendix*
\section{The spectral densities used in the calculations}

In this Appendix,  we collect the spectral densities corresponding to the $g_{\mu\nu}$ structure used in the calculations:
\begin{widetext}
\begin{eqnarray}
\rho^{pert} (s)& = &-  \frac{1}{3072 \pi^6}\int_0^1 dz \int_0^{1-z} dy\frac{1}{\varrho \kappa^8} \Bigg[y z \Big[(s y z (y + z - 1) -  m_c^2 \Big[(y^3 + y^2 (2 z - 1) +  2 y (z - 1) z  \nonumber \\
&+&(z - 1) z^2\Big]\Big]^2\Big[3 m_c^4 \Big[y^3 + y^2 (2 z - 1) 
+ 2 y (z - 1) z + (z - 1) z^2\Big]^2-26 m_c^2 s y z \Big[y^4  \nonumber\\
&+& y^3 (3 z - 2) + y^2 (4 z^2 - 5 z + 1)+y z (3 z^2 -  5 z + 2)+(z - 1)^2 z^2\Big] + 35 s^2 y^2 z^2  \nonumber \\
&\times& (y + z-1)^2\Bigg] \Theta[L(s, z, y)], \nonumber\\
\end{eqnarray}

\begin{eqnarray}
\rho^{\langle q\bar{q} \rangle_{\rho_N}}(s) & = &\frac{1}{4\pi^4}\int_0^1 dz \int_0^{1-z} dy \Bigg\{ \frac{m_c (y + z)}{4\kappa^5} \Bigg[3 m_c^4  
 \Big[y^3 + y^2 (2 z - 1) +  2 y (z - 1) z  \nonumber \\
&+& (z - 1) z^2\Big]^2 - 10 m_c^2 s y z \Big[y^4 + y^3 (3 z - 2) 
+ y^2 (4 z^2 - 5 z + 1) + y z (3 z^2 - 5 z + 2) \nonumber \\
&+& (z - 1)^2 z^2\Big] + 7 s^2 y^2 z^2 (y + z - 1)^2\Bigg]\langle\bar{u}u\rangle_{\rho_N} 
+  \frac{m_q m_c \sqrt{s_0^*} y z \varrho (y + z) }{\kappa^5} \Bigg[ 5 m_c^2 (y^3 \nonumber \\
&+& y^2 (2 z-1)+2 y (z-1) z+(z-1) z^2) 
-7 s y z (y+z-1)\Bigg]\langle u^{\dagger}u\rangle_{\rho_N} \Bigg\} \Theta[L(s, z, y)],
\end{eqnarray}

\begin{eqnarray}
\rho^{\langle GG \rangle_{\rho_N}}(s) & = &\frac{1}{96\pi^4}\int_0^1 dz \int_0^{1-z} dy \Bigg\{ \Bigg[-i \frac{m_q m_c}{\kappa^4}\Big[m_c^2 \Big(4 y^6 z
+ y^5 (24 z^2 - 8 z + 1) +  y^4 (52 z^3 - 44 z^2  \nonumber\\
&+&  6 z - 1) + y^3 z (60 z^3 - 80 z^2 + 55 z - 2) + y^2 z^2 (40 z^3 - 68 z^2 +   95 z - 34) \nonumber \\
&+& 6 y z^3 (2 z^3 - 4 z^2 + 13 z - 11)
+ 33 (z - 1) z^4\Big) + y z (y + z - 1) \Big(4 s_0^* \big(4 y^3 z \nonumber\\
&+& y^2 (16 z^2 - 4 z + 1) + 12 y (z - 1) z^2 + z^2\big) 
- 3 s \Big(4 y^3 z + y^2 (16 z^2 - 4 z + 1)  \nonumber\\
&+& 12 y (z - 1) z^2+ 17 z^2\Big)\Big) \Big]\Bigg]]\langle u^{\dagger}iD_0u\rangle_{\rho_N} + \frac{1}{1536(y-1)\varrho^2 \kappa^6}  \Bigg[m_c^4 (y-1) \Big(y^2+y (z-1)\nonumber\\
&+&(z-1) z\Big)^2 \Big(24 y^8 z+12 y^7 (11 z^2-8 z+3)
+ 3 y^6 (112 z^3-156 z^2+45 z-36) \nonumber \\
&+&2 y^5 (258 z^4-516 z^3+332 z^2-201 z+54)
+ y^4 (516 z^5-1320 z^4+1574 z^3-1198 z^2\nonumber\\
&+&471 z-36)+y^3 z (336 z^5-1032 z^4+1707 z^3 
-1769 z^2+858 z-132)+y^2 z^2 (132 z^5-468 z^4\nonumber\\
&+&1122 z^3-1253 z^2+807 z-180)+4 y z^3 (6 z^5
-24 z^4+152 z^3-155 z^2+96 z-27)+8 z^4 (23 z^3\nonumber\\
&-&29 z^2+9 z-3)\Big)-m_q^2 s (y-1) y z \Big(24 y^{10} z 
+12 y^9 (13 z^2-12 z+3)+y^8 (492 z^3-852 z^2 \nonumber\\
&-&89 z-180)+y^7 (984 z^4-2424 z^3+283 z^2
+672 z+360)+y^6 (1368 z^5-4296 z^4+2149 z^3 \nonumber \\
&+&1121 z^2-270 z-360)+y^5 (1368 z^6-5160 z^5
+4057 z^4-420 z^3+275 z^2-676 z+180) \nonumber \\
&+&y^4 (984 z^7-4296 z^6+4151 z^5-809 z^4 
+1804 z^3-1901 z^2+615 z-36)+y^3 z (492 z^7 \nonumber \\
&-&2424 z^6+2485 z^5+834 z^4+43 z^3-2396 z^2 
+1098 z-132)+y^2 z^2 (156 z^7-852 z^6+898 z^5 \nonumber \\
&+&1887 z^4-2785 z^3-27 z^2+903 z-180) 
+12 y (z-1)^2 z^3 (2 z^5-8 z^4+18 z^3+89 z^2 \nonumber \\
&-&12 z-9)+8 (z-1)^3 z^4 (23 z^2+30 z+3)\Big) 
-12 s y z^2 (y+z-1)^3 \Big(y^5 ((17 s+4) z-4) \nonumber \\
&+&y^4 (5 (3 s+2) z^2-2 (17 s+4) z+2)
+y^3 z (s (32 z^2-13 z+17)+4 (3 z^2-4 z+1)) \nonumber \\
&+&2 y^2 z^2 (s (16 z^2-32 z-1)+7 z^2-10 z+3)
-8 y (z-1) z^3 (4 s-z+1) \nonumber \\
&+&2 y^6+4 (z-1)^2 z^4\Big)\Bigg]\Big\langle \frac{\alpha_s}{\pi}G^2\Big\rangle_{\rho_N}\Bigg\} \Theta[L(s, z, y)], \nonumber \\
\end{eqnarray}

\begin{eqnarray}
\rho^{\langle qG\bar{q} \rangle_{\rho_N}}(s) & = &\frac{1}{96\pi^4}\int_0^1 dz \int_0^{1-z} dy \Bigg\{ \Bigg[-\frac{m_c}{\kappa^6} 
\times\Big[m_c^2 \Big(4 y^6 z + y^5 (24 z^2 - 8 z + 1) \nonumber \\
&+&  y^4 (52 z^3 - 44 z^2 - 2 z - 1) + y^3 z (60 z^3 
- 80 z^2 + 31 z + 6) +  y^2 z^2 (40 z^3 - 68 z^2 \nonumber \\
&+& 71 z - 18) + 2 y z^3 (6 z^3 - 12 z^2 + 35 z - 29) 
+ 33 (z - 1) z^4\Big) \Big(y^2 + y (z - 1) + (z - 1) z\Big)^3 \nonumber \\
&+& y z (y + z - 1) \Big(4 s_0^* (4 y^9 z + y^8 (28 z^2 - 16 z + 1)
+ 3 y^7 (28 z^3 - 36 z^2 + 25 z - 1)  \nonumber \\
&+&  y^6 (160 z^4 - 300 z^3 + 371 z^2 - 169 z + 3) +  y^5 (208 z^5 - 508 z^4 + 838 z^3 - 694 z^2 \nonumber \\
&+&  157 z - 1) + y^4 z (192 z^5 - 564 z^4 + 1116 z^3 
- 1276 z^2 + 567 z -  51) + 2 y^3 z^2 (62 z^5 - 210 z^4 \nonumber \\
&+& 463 z^3 - 642 z^2 + 409 z - 82) + y^2 (z - 1)^2 z^3 (52 z^3 - 92 z^2 + 239 z - 164) \nonumber \\
&+ &   3 y (z - 1)^3 z^4 (4 z^2 - 4 z + 17)
+ (z - 1)^3 z^5) - s \big(12 y^9 z + y^8 (84 z^2 - 48 z + 3) \nonumber \\
&+& y^7 (252 z^3 - 324 z^2 + 121 z - 9) + y^6 (480 z^4 
- 900 z^3 + 713 z^2 - 195 z + 9) + y^5 (624 z^5  \nonumber \\
&-& 1524 z^4 + 1730 z^3 - 986 z^2 + 159 z - 3) + y^4 z (576 z^5 - 1692 z^4 +2500 z^3 - 2060 z^2 \nonumber \\
&+&  709 z - 49) + 2 y^3 z^2 (186 z^5 - 630 z^4
+ 1125 z^3 - 1218 z^2 + 635 z - 98) + y^2 (z - 1)^2\nonumber \\
&\times&z^3 (156 z^3 - 276 z^2 + 589 z - 292) + y (z - 1)^3 z^4 (36 z^2 -36 z + 193) \nonumber \\
&+&  51 (z - 1)^3 z^5\big)\Big)\Big]
\Bigg]\langle \bar{u}iD_0iD_0u\rangle_{\rho_N}
+ \frac{m_c}{4 \kappa^7} \Bigg[ y z (y+z-1) \Big(2 s_0^* (4 y^9 z+y^8 (28 z^2 \nonumber \\
&-&16 z+1)+3 y^7 (28 z^3-36 z^2+25 z-1)
+ y^6 (160 z^4-300 z^3+371 z^2-169 m_cz+3) \nonumber \\
&+&y^5 (208 z^5-508 z^4+838 z^3-694 z^2+157 z-1) 
+ y^4 z (192 z^5-564 z^4+1116 z^3-1276 z^2 \nonumber \\
&+&567 z - 51)+2 y^3 z^2 (62 z^5-210 z^4+463 z^3-642 z^2 + 409 z-82) \nonumber \\
&+& y^2 (z-1)^2 z^3 (52 z^3-92 z^2+239 z
-164)+3 y (z-1)^3 z^4 (4 z^2-4 z+17)+(z-1)^3 z^5) \nonumber \\
&+& s (4 y^9 z+y^8 (28 z^2-16 z-59)+y^7 (84 z^3-108 z^2
- 401 z+177)+y^6 (160 z^4-300 z^3 \nonumber \\
&-&1197 z^2+1259 z-177)+y^5 (208 z^5-508 z^4-2022 z^3+3534 z^2 -1271 z+59) \nonumber \\
&+& y^4 z (192 z^5-564 z^4-2204 z^3
+ 5368 z^2-3185 z+425)+2 y^3 z^2 (62 z^5-210 z^4 \nonumber \\
&-& 751 z^3+2380 z^2-1945 z+464)+y^2 (z-1)^2 z^3 (52 z^3-92 z^2-861 z+760) \nonumber \\ 
&+&y (z-1)^3 z^4(12 z^2-12 z-173)+25 (z-1)^3 z^5)\Big)-m_c^2 (y^2+y (z-1) \nonumber \\
&+&(z-1) z)^3 \Big(4 y^6 z+y^5 (24 z^2-8 z
-35)+y^4 (52 z^3-44 z^2-170 z+35)+y^3 z (60 z^3 \nonumber \\
&-&80 z^2-353 z+174)+y^2 z^2 (40 z^3-68 z^2-301 z 
+234)+2 y z^3 (6 z^3-12 z^2-37 z+43) \nonumber \\
&+&9 (z-1)z^4 \Big) \Bigg]\langle \bar{u}g_s\sigma Gu\rangle_{\rho_N}\Bigg\} \Theta[L(s, z, y)], \nonumber \\
\end{eqnarray}
where 
\begin{eqnarray}
\varrho&=& (y+z-1), \nonumber \\
\kappa&=&(y^2+y (z-1)+(z-1) z), \nonumber \\
L[s, y,  z] &=& \frac{((-1 + y)(-(syz(-1 + y + z)) + 
       mc^2 (y^3 + 2 y (-1 + z) z + (-1 + z) z^2 + 
          y^2 (-1 + 2z))))}{(y^2 + y (-1 + z) + (-1 + z) z)^2},
\end{eqnarray}
\end{widetext}
and $\Theta$ is the usual step function.


%


\begin{thebibliography}{99}

\bibitem{Choi:2003ue} 
  S.~K.~Choi {\it et al.} [Belle Collaboration],
  ``Observation of a narrow charmonium - like state in exclusive $B^{+} \rightarrow K^{+}\pi^{+}\pi^{-}J/\psi$ decays,''
  Phys.\ Rev.\ Lett.\  {\bf 91}, 262001 (2003)
  [hep-ex/0309032].

\bibitem{Acosta:2003zx} 
  D.~Acosta {\it et al.} [CDF Collaboration],
  ``Observation of the narrow state $X(3872) \to J/\psi \pi^+ \pi^-$ in $\bar{p}p$ collisions at $\sqrt{s} = 1.96$ TeV,''
  Phys.\ Rev.\ Lett.\  {\bf 93}, 072001 (2004)
  [hep-ex/0312021].
  
\bibitem{Abazov:2004kp} 
  V.~M.~Abazov {\it et al.} [D0 Collaboration],
  ``Observation and properties of the $X(3872)$ decaying to $J/\psi \pi^+ \pi^-$ in $p\bar{p}$ collisions at $\sqrt{s} = 1.96$ TeV,''
  Phys.\ Rev.\ Lett.\  {\bf 93}, 162002 (2004)
  [hep-ex/0405004].
  
\bibitem{Aubert:2004ns} 
  B.~Aubert {\it et al.} [BaBar Collaboration],
  ``Study of the $B \to J/\psi K^- \pi^+ \pi^-$ decay and measurement of the $B \to X(3872) K^-$ branching fraction,''
  Phys.\ Rev.\ D {\bf 71}, 071103 (2005)
  [hep-ex/0406022].

\bibitem{Abe:2005iya} 
  K.~Abe {\it et al.} [Belle Collaboration],
``Experimental constraints on the possible $ J^{PC} $ quantum numbers of the X(3872),''
  hep-ex/0505038.
  
\bibitem{Nielsen:2009uh} 
  M.~Nielsen, F.~S.~Navarra and S.~H.~Lee,
  ``New Charmonium States in QCD Sum Rules: A Concise Review,''
  Phys.\ Rept.\  {\bf 497}, 41 (2010)
  [arXiv:0911.1958 [hep-ph]].
  
\bibitem{Choi:2011fc} 
  S.-K.~Choi {\it et al.},
  ``Bounds on the width, mass difference and other properties of $X(3872) \rightarrow pi+pi-J/psi$ decays,''
  Phys.\ Rev.\ D {\bf 84}, 052004 (2011)
  [arXiv:1107.0163 [hep-ex]].
  
\bibitem{Ortega:2017hpw} 
  P.~G.~Ortega, D.~R.~Entem, F.~Fernandez and E.~Ruiz Arriola,
  ``Counting states and the Hadron Resonance Gas: Does X(3872) count?,''
  arXiv:1707.01915 [hep-ph].
  
\bibitem{Yu:2017bsj} 
  G.~L.~Yu, Z.~G.~Wang and Z.~Y.~Li,
  ``The analysis of the charmonium-like states $X^{*}(3860)$,$X(3872)$, $X(3915)$, $X(3930)$ and $X(3940)$ according to its strong decay behaviors,''
  arXiv:1704.06763 [hep-ph].
  
\bibitem{Zhou:2017dwj} 
  Z.~Y.~Zhou and Z.~Xiao,
  ``Understanding $X(3862)$, $X(3872)$, and $X(3930)$ in a Friedrichs-model-like scheme,''
  Phys.\ Rev.\ D {\bf 96}, no. 5, 054031 (2017)
  [arXiv:1704.04438 [hep-ph]].
  
\bibitem{Wang:2015rcz} 
  W.~Wang and Q.~Zhao,
  ``Decipher the short-distance component of $X(3872)$ in $B_c$ decays,''
  Phys.\ Lett.\ B {\bf 755}, 261 (2016)
  [arXiv:1512.03123 [hep-ph]].
  
\bibitem{Ferretti:2015fba} 
  J.~Ferretti, E.~Santopinto and H.~Garcia-Tecocoatzi,
  ``Higher charmonia and bottomonia. Nature of the X(3872),''
  arXiv:1510.00433 [hep-ph].
  
\bibitem{Pena:2014pea} 
  C.~Pena and L.~Jacak,
  ``A braid model for the particle X(3872),''
  arXiv:1411.1574 [hep-ph].
  
\bibitem{Guo:2014hqa} 
  F.~K.~Guo, C.~Hidalgo-Duque, J.~Nieves, A.~Ozpineci and M.~P.~Valderrama,
  ``Detecting the long-distance structure of the $X$(3872),''
  Eur.\ Phys.\ J.\ C {\bf 74}, no. 5, 2885 (2014)
  [arXiv:1404.1776 [hep-ph]].
  
\bibitem{Wang:2013vex} 
  Z.~G.~Wang and T.~Huang,
  ``Analysis of the $X(3872)$, $Z_c(3900)$ and $Z_c(3885)$ as axial-vector tetraquark states with QCD sum rules,''
  Phys.\ Rev.\ D {\bf 89}, no. 5, 054019 (2014)
  [arXiv:1310.2422 [hep-ph]].
  
\bibitem{Baru:2013rta} 
  V.~Baru, E.~Epelbaum, A.~A.~Filin, C.~Hanhart, U.-G.~Meissner and A.~V.~Nefediev,
  ``Quark mass dependence of the X(3872) binding energy,''
  Phys.\ Lett.\ B {\bf 726}, 537 (2013)
  [arXiv:1306.4108 [hep-ph]].
  
\bibitem{Wang:2013kva} 
  P.~Wang and X.~G.~Wang,
  ``Study on X(3872) from effective field theory with pion exchange interaction,''
  Phys.\ Rev.\ Lett.\  {\bf 111}, no. 4, 042002 (2013)
  [arXiv:1304.0846 [hep-ph]].
  
\bibitem{Cui:2011be} 
  C.~Y.~Cui, Y.~L.~Liu, G.~B.~Zhang and M.~Q.~Huang,
  ``Can X(3872) be a $J^{P}=2^{-}$ tetraquark state?,''
  Commun.\ Theor.\ Phys.\  {\bf 57}, 1033 (2012)
  [arXiv:1112.5976 [hep-ph]].
    
\bibitem{Takizawa:2012hy} 
  M.~Takizawa and S.~Takeuchi,
  ``X(3872) as a hybrid state of charmonium and the hadronic molecule,''
  PTEP {\bf 2013}, 093D01 (2013)
  [arXiv:1206.4877 [hep-ph]].
  
\bibitem{Coito:2012vf} 
  S.~Coito, G.~Rupp and E.~van Beveren,
  ``X(3872) is not a true molecule,''
  Eur.\ Phys.\ J.\ C {\bf 73}, no. 3, 2351 (2013)
  [arXiv:1212.0648 [hep-ph]].
  
\bibitem{Ferretti:2014xqa} 
  J.~Ferretti, G.~Galata and E.~Santopinto,
  ``Quark structure of the $X(3872)$ and $\chi_b(3P)$ resonances,''
  Phys.\ Rev.\ D {\bf 90}, no. 5, 054010 (2014)
  [arXiv:1401.4431 [nucl-th]].
  
\bibitem{Karliner:2014lta} 
  M.~Karliner and J.~L.~Rosner,
  ``$X(3872)$, $X_b$, and the $\chi_{b1}(3P)$ state,''
  Phys.\ Rev.\ D {\bf 91}, no. 1, 014014 (2015)
  [arXiv:1410.7729 [hep-ph]].
 
\bibitem{Achasov:2015oia} 
  N.~N.~Achasov and E.~V.~Rogozina,
  ``$X(3872), I^G(J^{PC})=0^+(1^{++})$, as the $\chi_{1c}(2P)$ charmonium,''
  Mod.\ Phys.\ Lett.\ A {\bf 30}, no. 33, 1550181 (2015)
  [arXiv:1501.03583 [hep-ph]].
  
\bibitem{Larionov:2015nea} 
  A.~B.~Larionov, M.~Strikman and M.~Bleicher,
  ``Determination of the structure of the $X(3872)$ in $\bar p A$ collisions,''
  Phys.\ Lett.\ B {\bf 749}, 35 (2015)
  [arXiv:1502.03311 [nucl-th]].
  
\bibitem{Larionov:2015kxn} 
  A.~Larionov, M.~Strikman and M.~Bleicher,
  ``Test of the $X(3872)$ Structure in Antiproton-Nucleus Collisions,''
  AIP Conf.\ Proc.\  {\bf 1735}, 060005 (2016)
  [arXiv:1512.05536 [nucl-th]].
  
\bibitem{Meng:2014ota} 
  C.~Meng, J.~J.~Sanz-Cillero, M.~Shi, D.~L.~Yao and H.~Q.~Zheng,
  ``Refined analysis on the X(3872) resonance,''
  Phys.\ Rev.\ D {\bf 92}, no. 3, 034020 (2015)
  [arXiv:1411.3106 [hep-ph]].
  
\bibitem{Chen:2013pya} 
  W.~Chen, H.~y.~Jin, R.~T.~Kleiv, T.~G.~Steele, M.~Wang and Q.~Xu,
  ``QCD sum-rule interpretation of X(3872) with $J^{PC}=1^{++}$ mixtures of hybrid charmonium and $\overline{D}D^*$ molecular currents,''
  Phys.\ Rev.\ D {\bf 88}, no. 4, 045027 (2013)
  [arXiv:1305.0244 [hep-ph]].
  
\bibitem{Maiani:2004vq} 
  L.~Maiani, F.~Piccinini, A.~D.~Polosa and V.~Riquer,
  ``Diquark-antidiquarks with hidden or open charm and the nature of X(3872),''
  Phys.\ Rev.\ D {\bf 71}, 014028 (2005)
  [hep-ph/0412098].
  
\bibitem{Bugg:2004rk} 
  D.~V.~Bugg,
 ``Reinterpreting several narrow `resonances' as threshold cusps,''
  Phys.\ Lett.\ B {\bf 598}, 8 (2004)
  [hep-ph/0406293].
  
\bibitem{Li:2004sta} 
  B.~A.~Li,
  ``Is X(3872) a possible candidate of hybrid meson,''
  Phys.\ Lett.\ B {\bf 605}, 306 (2005)
  [hep-ph/0410264].
  
\bibitem{Seth:2004zb} 
  K.~K.~Seth,
  ``An Alternative Interpretation of X(3872),''
  Phys.\ Lett.\ B {\bf 612}, 1 (2005)
  [hep-ph/0411122].
  
\bibitem{Swanson:2003tb} 
  E.~S.~Swanson,
  ``Short range structure in the X(3872),''
  Phys.\ Lett.\ B {\bf 588}, 189 (2004)
  [hep-ph/0311229].
  
\bibitem{Wang:2013daa} 
  Z.~G.~Wang and T.~Huang,
  ``Possible assignments of the $X(3872)$, $Z_c(3900)$ and $Z_b(10610)$ as axial-vector molecular states,''
  Eur.\ Phys.\ J.\ C {\bf 74}, no. 5, 2891 (2014)
  [arXiv:1312.7489 [hep-ph]].

\bibitem{Braaten:2013poa} 
  E.~Braaten and D.~Kang,
  ``$J/{\psi}$ ${\omega}$ Decay Channel of the $X$(3872) Charm Meson Molecule,''
  Phys.\ Rev.\ D {\bf 88}, no. 1, 014028 (2013)
  [arXiv:1305.5564 [hep-ph]].
  
\bibitem{Chen:2016qju} 
  H.~X.~Chen, W.~Chen, X.~Liu and S.~L.~Zhu,
  ``The hidden-charm pentaquark and tetraquark states,''
  Phys.\ Rept.\  {\bf 639}, 1 (2016)
  [arXiv:1601.02092 [hep-ph]].
 
  
  \bibitem{Matheus:2006xi} 
  R.~D.~Matheus, S.~Narison, M.~Nielsen and J.~M.~Richard,
  ``Can the X(3872) be a $1^{++}$ four-quark state?,''
  Phys.\ Rev.\ D {\bf 75}, 014005 (2007)
  [hep-ph/0608297].
  
\bibitem{Kang:2016jxw} 
  X.~W.~Kang and J.~A.~Oller,
  ``Different pole structures in line shapes of the $X(3872)$,''
  Eur.\ Phys.\ J.\ C {\bf 77}, no. 6, 399 (2017)
  [arXiv:1612.08420 [hep-ph]].
  
\bibitem{Sundu:2016oda} 
  H.~Sundu,
  ``The Mass and Current-Meson Coupling Constant of the Exotic X(3872) State from QCD Sum Rules,''
  SDU J.\ Nat.\ Appl.\ Sci.\  {\bf 20}, no. 3, 448 (2016).
  
\bibitem{Veliev:2017fpa} 
  E.~V.~Veliev, S.~Gunaydin and H.~Sundu,
  ``Thermal properties of exotic $X(3872)$ state via QCD sum rule,''
  arXiv:1707.03714 [hep-ph].

\bibitem{Giacosa:2015nwa} 
  F.~Giacosa,
  ``Non-conventional mesons at PANDA,''
  J.\ Phys.\ Conf.\ Ser.\  {\bf 599}, no. 1, 012004 (2015)
  [arXiv:1502.02682 [hep-ph]].
  
\bibitem{Prencipe:2015cgg} 
  E.~Prencipe {\it et al.} [PANDA Collaboration],
  AIP Conf.\ Proc.\  {\bf 1735}, 060011 (2016)
  [arXiv:1512.05496 [hep-ex]].
  
\bibitem{Biswas:2015paa} 
  S.~Biswas {\it et al.},
  ``Measurement of the spark probability of a GEM detector for the CBM muon chamber (MuCh),''
  Nucl.\ Instrum.\ Meth.\ A {\bf 800}, 93 (2015)
  [arXiv:1504.00001 [physics.ins-det]].
  
 \bibitem{fair} 	http://www.gsi.de/fair/experiments/CBM/index.e.html
 
\bibitem{panda} http://www-panda.gsi.de/auto/phy/home.htm
 
 \bibitem{Friman} B. Friman et al., The CBM Physics Book: Compressed Baryonic Matter in Laboratory Experiments, Springer, Heidelberg (2011).
 
\bibitem{Lutz:2009ff} 
  M.~F.~M.~Lutz {\it et al.} [PANDA Collaboration],
  ``Physics Performance Report for PANDA: Strong Interaction Studies with Antiprotons,''
  arXiv:0903.3905 [hep-ex].
 
  \bibitem{nika} 	http://nica.jinr.ru/innovation.php  
  
\bibitem{Thomas:2007gx} 
  R.~Thomas, T.~Hilger and B.~Kampfer,
  ``Four-quark condensates in nucleon QCD sum rules,''
  Nucl.\ Phys.\ A {\bf 795}, 19 (2007)
  [arXiv:0704.3004 [hep-ph]].
  
\bibitem{Cohen:1994wm} 
  T.~D.~Cohen, R.~J.~Furnstahl, D.~K.~Griegel and X.~m.~Jin,
  ``QCD sum rules and applications to nuclear physics,''
  Prog.\ Part.\ Nucl.\ Phys.\  {\bf 35}, 221 (1995)
  [hep-ph/9503315].


\bibitem{Serot:1984ey} 
  B.~D.~Serot and J.~D.~Walecka,
  ``The Relativistic Nuclear Many Body Problem,''
  Adv.\ Nucl.\ Phys.\  {\bf 16}, 1 (1986).
  
\bibitem{Azizi:2014yea} 
  K.~Azizi and N.~Er,
  ``Properties of nucleon in nuclear matter: once more,''
  Eur.\ Phys.\ J.\ C {\bf 74}, 2904 (2014)
  [arXiv:1401.1680 [hep-ph]].
  
\bibitem{Jin:1992id}
  X.~m.~Jin, T.~D.~Cohen, R.~J.~Furnstahl and D.~K.~Griegel,
  ``QCD sum rules for nucleons in nuclear matter. 2.,''
  Phys.\ Rev.\ C {\bf 47}, 2882 (1993).

\bibitem{Agaev:2016srl} 
  S.~S.~Agaev, K.~Azizi and H.~Sundu,
  ``Application of the QCD light cone sum rule to tetraquarks: the strong vertices $X_bX_b\rho$ and $X_cX_c\rho$,''
  Phys.\ Rev.\ D {\bf 93}, no. 11, 114036 (2016)
  [arXiv:1605.02496 [hep-ph]].



\bibitem{Cohen:1991nk} 
  T.~D.~Cohen, R.~J.~Furnstahl and D.~K.~Griegel,
  ``Quark and gluon condensates in nuclear matter,''
  Phys.\ Rev.\ C {\bf 45}, 1881 (1992).
  
\bibitem{Belyaev:1982cd} 
  V.~M.~Belyaev and B.~L.~Ioffe,
  ``Determination of the baryon mass and baryon resonances from the quantum-chromodynamics sum rule. Strange baryons,''
  Sov.\ Phys.\ JETP {\bf 57}, 716 (1983)
  [Zh.\ Eksp.\ Teor.\ Fiz.\  {\bf 84}, 1236 (1983)].
  
\bibitem{Ioffe:2005ym} 
  B.~L.~Ioffe,
  ``QCD at low energies,''
  Prog.\ Part.\ Nucl.\ Phys.\  {\bf 56}, 232 (2006)
  [hep-ph/0502148].
 
  
\bibitem{Azizi:2016dmr} 
  K.~Azizi, N.~Er and H.~Sundu,
 ``Scalar and vector self-energies of heavy baryons in nuclear medium,''
  Nucl.\ Phys.\ A {\bf 960}, 147 (2017)
  Erratum: [Nucl.\ Phys.\ A {\bf 962}, 122 (2017)]
  [arXiv:1605.05535 [hep-ph]].
  
\bibitem{Hosaka:2016ypm} 
  A.~Hosaka, T.~Hyodo, K.~Sudoh, Y.~Yamaguchi and S.~Yasui,
 ``Heavy Hadrons in Nuclear Matter,''
  Prog.\ Part.\ Nucl.\ Phys.\  {\bf 96}, 88 (2017)
  [arXiv:1606.08685 [hep-ph]].

\bibitem{Wang:2012xk} 
  Z.~G.~Wang,
  ``Analysis of the doubly heavy baryons in the nuclear matter with the QCD sum rules,''
  Eur.\ Phys.\ J.\ C {\bf 72}, 2099 (2012)
  [arXiv:1205.0605 [hep-ph]].
  
\bibitem{Wang:2011yj} 
  Z.~G.~Wang,
  ``Analysis of the $\Sigma_Q$ baryons in the nuclear matter with the QCD sum rules,''
  Phys.\ Rev.\ C {\bf 85}, 045204 (2012)
  [arXiv:1109.2180 [hep-ph]].

\bibitem{Wang:2011hta} 
  Z.~G.~Wang,
  ``Analysis of the $\Lambda_Q$ baryons in the nuclear matter with the QCD sum rules,''
  Eur.\ Phys.\ J.\ C {\bf 71}, 1816 (2011)
  [arXiv:1108.4251 [hep-ph]].
  
\bibitem{Furnstahl:1995nd} 
  R.~J.~Furnstahl, X.~m.~Jin and D.~B.~Leinweber,
  ``New QCD sum rules for nucleons in nuclear matter,''
  Phys.\ Lett.\ B {\bf 387}, 253 (1996)
  [nucl-th/9511007].

\bibitem{Drukarev:2004fn} 
  E.~G.~Drukarev, M.~G.~Ryskin and V.~A.~Sadovnikova,
  ``QCD sum rules description of nucleons in asymmetric nuclear matter,''
  Phys.\ Rev.\ C {\bf 70}, 065206 (2004)
  [nucl-th/0406027].
  
\bibitem{Drukarev:2001wd} 
  E.~G.~Drukarev, M.~G.~Ryskin and V.~A.~Sadovnikova,
  ``QCD condensates and hadron parameters in nuclear matter: Selfconsistent treatment, sum rules and all that,''
  Prog.\ Part.\ Nucl.\ Phys.\  {\bf 47}, 73 (2001)
  [nucl-th/0106049].

\bibitem{Koike:1996ga} 
  Y.~Koike and A.~Hayashigaki,
  ``QCD sum rules for rho, omega, phi meson - nucleon scattering lengths and the mass shifts in nuclear medium,''
  Prog.\ Theor.\ Phys.\  {\bf 98}, 631 (1997)
  [nucl-th/9609001].
  
\bibitem{Hayashigaki:2000es} 
  A.~Hayashigaki,
  Phys.\ Lett.\ B {\bf 487}, 96 (2000)
  doi:10.1016/S0370-2693(00)00760-7
  [nucl-th/0001051].
  
\bibitem{Ohtani:2016pyk} 
  K.~Ohtani, P.~Gubler and M.~Oka,
  ``Negative-parity nucleon excited state in nuclear matter,''
  Phys.\ Rev.\ C {\bf 94}, no. 4, 045203 (2016)
  [arXiv:1606.09434 [hep-ph]].
  
\bibitem{Suzuki:2015est} 
  K.~Suzuki, P.~Gubler and M.~Oka,
  ``D meson mass increase by restoration of chiral symmetry in nuclear matter,''
  Phys.\ Rev.\ C {\bf 93}, no. 4, 045209 (2016)
  [arXiv:1511.04513 [hep-ph]].

\end{thebibliography}
\end{document}